\documentstyle[aps,preprint]{revtex}
\begin{document}
\draft

\preprint{KIAS-P98009, SNUTP98-064}

\title{Thermodynamics of doubly charged CGHS model and $D1-D5-KK$ black
holes of IIB supergravity}
\author{Youngjai Kiem$^{(a)}$\footnote{ykiem@kiasph.kaist.ac.kr}, 
        Chang-Yeong Lee$^{(b)}$\footnote{leecy@phy.sejong.ac.kr}, 
    and Dahl Park$^{(c)}$\footnote{dpark@ctp.snu.ac.kr} }
\address{ $^{(a)}$ School of Physics, KIAS, Seoul 130-012, Korea\\
          $^{(b)}$ Department of Physics, Sejong University, 
                   Seoul 133-747, Korea\\
          $^{(c)}$ Center for Theoretical Physics, Seoul National
                   University, Seoul 151-742, Korea}
\maketitle

\begin{abstract}
We study the doubly charged Callan-Giddings-Harvey-Strominger
(CGHS) model, which has black hole solutions that were found to 
be $U$-dual to the $D1$-$D5$-$KK$ black holes of the IIB 
supergravity.  We derive the action of the model via a 
spontaneous compactification on $S^3$ of the IIB supergravity  
on $S^1 \times T^4$ and obtain the general static solutions
including black holes corresponding to certain non-asymptotically 
flat black holes in the IIB supergravity.  Thermodynamics of them
is established by computing the entropy, temperature, chemical 
potentials, and mass in the two-dimensional setup, and the first 
law of thermodynamics is explicitly verified.  The entropy is in 
precise agreement with that of the $D1-D5-KK$ black holes, and the 
mass turns out to be consistent with the infinite Lorentz boost 
along the $M$ theory circle that is a part of the aforementioned 
$U$-dual chain.
\end{abstract}

%\pacs{04.65.+e, 04.60.Kz, 11.25.Sq, 04.70.Dy}

%\newpage

\section{Introduction}

The discrete light-cone quantization is increasingly
becoming of importance as we gain more understanding of the
matrix theory formulation of $M$ theory \cite{dlcq}.  If we lift 
the ten-dimensional type IIA $D$-particle solutions into
the eleven dimensions and take the
infinite Lorentz boost along the $M$ theory circle {\em a la} 
Seiberg \cite{seiberg}, we get
the description of the gravitons traveling along a discrete
light-cone \cite{hks}.  The same consideration for the $M$ theory 
compactified on a $p$-torus with $p=2,3,4$ yields an Anti de Sitter
space-time tensored with a sphere \cite{hk}.  In the case of $p=5$, 
the main object of interest is $D1$-$D5$ bound states \cite{seiberg}.  
In this context,
the infinitely boosted version of the $D1$-$D5$-$KK$ black holes
of the IIB supergravity on $S^1 \times T^5$ along the $M$ theory
circle is expected to play a role \cite{hk}.  From the point of view 
of the black hole physics, the same $D1$-$D5$-$KK$ black holes of the 
five-dimensional IIB supergravity is the prime example where we have 
an analytic handle over their microscopic quantum 
description \cite{vafa}.  
The dimensionally uplifted into eleven dimensions, 
infinitely boosted, and further dimensionally reduced versions
of the original $D1$-$D5$-$KK$ solutions, 
with appropriate $U$-dual transformations, are the non-asymptotically
flat solutions corresponding to the deletion of one in
the harmonic function produced by the fivebrane
charge \cite{hks}.  Thus, the space-time geometry is essentially
that of the near-horizon part of the $D1$-$D5$-$KK$ black holes
\cite{stro}. Upon further $S$-dual transformation, these
solutions become the doubly charged Callan-Giddings-Harvey-Strominger
(CGHS) black holes \cite{cghs}, establishing
a $U$-duality between black holes in differing space-time 
dimensions \cite{hyun}.

Even at the classical level, the analysis of the thermodynamics
for non-asymptotically flat solutions needs a careful treatment,
making the thermodynamic analysis of the non-asymptotically flat 
five-dimensional IIB black holes difficult.
However, the aforementioned $U$-dual relationship implies that
we can tackle the same problem in a much simpler setting 
of the two-dimensional dilaton gravity \cite{2dg} of the CGHS type, 
and that is what we achieve in
this paper.  In Section II, we start from spontaneously compactifying
further on $S^3$ the IIB supergravity bosonic action compactified 
on $S^1 \times T^4$.  When there is a non-vanishing magnetic field
produced by NS-fivebranes (D-fivebranes before the $S$-dual 
transformation) wrapped along the $S^1 \times T^4$, the spontaneous
compactification produces a two-dimensional cosmological constant
inversely proportional to the NS field strength.  The resulting
dilaton-gravity sector of the two-dimensional theory precisely 
becomes that of the CGHS model.  In the 
two-dimensional action, in addition, there are 
two $U(1)$ gauge fields produced by the fundemantal string wrapping
along the $S^1$ ($D$-string wrapping number before $S$-dual
transformation) and the Kaluza-Klein momentum along the same circle, 
as well as two scalar fields
representing the size of the circle and the four-torus.
The two-dimensional action shows the manifest $T$-dual invariance
when we take the $T$-dual transformation along the circle.
In Section III, we solve the static equations of motion of
the two dimensional model in a conformal gauge to get the
general static solutions including all black hole 
and naked singularity solutions.  These black holes carry
the aforementioned two $U(1)$ charges as well as mass,
and we demonstrate the no-scalar hair property for the 
scalar fields.  We note that the original fivebranes charge is 
traded to become the cosmological constant of the 
two-dimensional theory under 
the spontaneous compactification.  By transforming these solutions to
a radial gauge, we find that the black hole solutions of the 
two-dimensional theory are the non-asymptotically 
flat solutions of the 
IIB supergravity that were found to be $U$-dual to 
the $D1$-$D5$-$KK$ black hole solutions.  From the two-dimensional
point of view, however, they are asymptotically flat and the
usual thermodynamic analysis applies.  In Section IV, we 
explicitly evaluate the theromodynamic quantities of the
black holes such as
entropy, mass, temperature, and chemical potentials via the
two-dimensional analysis, and explicitly verify the first law of
thermodynamics.  For the entropy formula, we
find an agreement with the entropy of the $D1$-$D5$-$KK$ 
black holes.  This is consistent with the statement that
these two black holes are $U$-dual to each other, since
the entropy is a $U$-dual invariant quantity.  Similarly,
we find a (non-extremal) mass formula that is consistent 
with the infinite Lorentz boost along the $M$ theory circle
as prescribed by Seiberg \cite{seiberg} \cite{hks}.  We also comment 
on the particularly simple form of the mass formula in this section.  
In section V, we discuss our results and comment on a related model 
of McGuigan, Nappi and Yost \cite{mny}.  The doubly charged
CGHS model that we derive in this paper is an interesting model
theory of gravity whose black holes have a non-vanishing entropy
in the extremal limit.

\section{Lower dimensional setup of the problem}

In this section, we reformulate the problem of 
five-dimensional IIB black holes
in terms of the two-dimensional black holes.
Let us consider a IIB black hole on $M_5 \times S^1 
\times T^4$, where $D$ five-branes wrap the five-torus $S^1 \times T^4$
$Q_5$ times, $D$-strings wrap the circle $S^1$ $Q_2$ times, and we
have the Kaluza-Klein (KK) momentum $Q$ along the circle $S_1$.
Here $M_5$ is the non-compact five-dimensional space-time.
This black hole has been shown in Ref. \cite{hks} to be $U$-dual
to the doubly charged CGHS black hole.  The precise chain of the dual
transformations are as follows.  We take an appropriate number 
of $T$-duals to turn the $D$ five-branes into $D$-particles.  
We lift this solution
into eleven dimensions and take an infinite boost 
along the $M$-theory
circle following Seiberg \cite{seiberg}.  After dimensionally 
reducing it to ten dimensions
and taking $T$-duals back, we further take an $S$-dual transformation.  
In the resulting space time geometry, $M_5$ gets spontaneously
compactified to yield $M_2 \times S^3$ where the three-sphere $S^3$ has a 
constant radius and $M_2$ is a two dimensional manifold.  At the same time, 
the size of the $T^4$ becomes a constant.  Now, the $(2+1)$-dimensional part 
of the space-time $M_2 \times S^1$ turns out to be a black string solution,
which, upon the compactification along the $S^1$, becomes a doubly charged
CGHS black hole.  

Technically speaking, the aforementioned analysis was done at the
level of classical solutions.  What we want to do in this section is to
follow the same prescription at the level of the classical action.
The ten-dimensional IIB supergravity action with the non-vanishing
RR two-form gauge field $B$, graviton $g^{(10)}_{\mu \nu}$, and the dilaton 
$\phi^{(10)}$ is given by
\begin{equation}
I = \frac{1}{16 \pi (8\pi^6) } \int d^{10} x  
\sqrt{-g^{(10)}} \left( e^{-2 \phi^{(10)}} R^{(10)}
  + 4 e^{-2 \phi^{(10)}} (D \phi^{(10)} )^2 - \frac{1}{12} H^{(10) 2} \right) ,
\label{iibo}
\end{equation}
where the field strength $H^{(10)} =  d B$.  We use a unit where
$\alpha^{\prime} = 1$, the string coupling 
$g \equiv \exp ( \phi^{(10)} )$, and the ten-dimensional Newton's
constant $G_N^{10} = 8 \pi^6$.
After the taking an $S$-dual transformation, we get 
\begin{equation}
 I = \frac{1}{16 \pi (8 \pi^6) } 
   \int d^{10} x  \sqrt{-g^{(10)}} e^{2 \phi^{(10)}} \left( R^{(10)}
  + 4  (D \phi^{(10)} )^2 - \frac{1}{12} H^{(10) 2} \right)   ,
\label{iib}
\end{equation}
where the original RR two-form gauge field changes to the NS-NS two-form
gauge field.   We consider the case when $NS$ ($D$ before the $S$-duality)
five-branes are wrapped along the $S^1 \times T^4$, and 
we allow the fundamental string ($D$-string) wrapping and the momentum
along the circle $S^1$.
We thus choose the metric of the form 
\[ ds^2 =  g_{\mu \nu}^{(6)} dx^{\mu} dx^{\nu} + e^{\psi} dx^m dx^m \]
where the index $m$ ranges from six to nine and the six-dimensional metric
$g^{(6)}_{\mu \nu}$ is given by
\[ g^{(6)}_{\mu \nu} = \left( \begin{array}{cc} 
             g^{(5)}_{\alpha \beta} + e^{\psi_1} A_{\alpha} A_{\beta}  & 
             e^{\psi_1} A_{\alpha} \\
             e^{\psi_1} A_{\beta} & e^{\psi_1} \end{array} \right)  , \]
to get the five-dimensional action
\begin{eqnarray}
I=&&\frac{1}{4 \pi^2} 
\int d^5 x \sqrt{- g^{(5)} } e^{2 \phi_1} \left(
R^{(5)} + 4 ( D \phi_1 )^2 - (D \psi )^2  - \frac{1}{4} (D \psi_1 )^2
\right. \nonumber \\
&&~~~~~~~~~~~~~~~~~~~~~~~~~~~~~~~~~~~~~~~~~~~\left. - \frac{1}{12} 
H^{\prime 2}  - \frac{1}{4} e^{-\psi_1} F_2^2 -
 \frac{1}{4} e^{\psi_1} F^2 \right)    
\label{fived}
\end{eqnarray}
where the five-dimensional dilaton $2 \phi_1 = 2 \phi^{(10)} + 
2 \psi + \psi_1 /2 $.  The volume of the unit five-torus, $(2 \pi )^5$,
was multiplied to the numerical factor in front of the action.
The field strength $F = d A$ has the Kaluza-Klein momentum as its 
charge, the field strength $F_{2 \alpha \beta} = H_{\alpha \beta x^5}$ 
comes from the string wrapping along the $S^1$, and $H^{\prime}
= H - A \wedge F_2$ originates from the five-brane wrapping along
$S^1 \times T^4$.
The $s$-wave dynamics of the above five-dimensional system are summarized by
the following two dimensional action \cite{klp} 
\begin{equation}
 I = \frac{1}{2} \int d^2 x \sqrt{- g } e^{- 2 \phi } \Big(
R + 6 e^{2 \psi_2} + 4 ( D \phi )^2 - (D \psi )^2  - \frac{1}{4} 
(D \psi_1 )^2 -3 (D \psi_2 )^2 
\label{twod}
\end{equation}
\[ - \frac{1}{2} e^{6 \psi_2 } H_0^2  - \frac{1}{4} e^{-\psi_1} F_2^2 -
 \frac{1}{4} e^{\psi_1} F^2 \Big)   \]
where we set the five-dimensional metric as
\[ ds^2 = g_{\alpha \beta} dx^{\alpha} dx^{\beta} + e^{- 2 \psi_2 }
   d \Omega_{(3)}  \]
and we require that all fields are independent of the angular coordinates.
The volume of the unit three-sphere, $2 \pi^2$, was multiplied
to the numerical factor in front of the action.
Here, $d \Omega_{(3)}$ is the metric on the standard unit three-sphere
$S^3$ and $g_{\alpha \beta}$ is the two dimensional metric.  In fact,
$H^{\prime}_{ijk} = H_0 \epsilon_{ijk}$ and
$H_0$ is a zero-form field strength on the two-dimensional
space time, since $H^{\prime}$ is proportional to the volume form 
$\epsilon_{ijk}$ of the 
three-sphere.  As such, it satisfies the Bianchi identity 
$\partial_{\alpha} H_0 = 0$
which forces it to be a constant.  
This number is proportional to the five-brane
wrapping number along $S^1 \times T^4$.  The two dimensional
dilaton field $\phi$ is given by $2 \phi = - 2 \phi_1 + 3 \psi_2$.

For a generic non-degerate $\psi_2$, i.e., $d \psi_2 \ne 0$, the action
Eq. (\ref{twod}) simply describes the bosonic $s$-wave sector of the 
five non-compact dimensional IIB supergravity.  
Our main interest in this paper 
however is to concentrate on the degenerate case 
($d \psi_2 = 0$) where we have 
a spontaneous compactification on the three-sphere $S^3$ of the
five-manifold, as explained
earlier in this section.  For this purpose, we consider the
equation of motion for $\psi_2$ field   
\begin{equation}
2 e^{2 \phi} D^{\alpha} ( e^{-2 \phi} D_{\alpha} \psi_2 )
 + 4 e^{2 \psi_2} - H_0^2 e^{6 \psi_2} = 0 . 
\label{phi3}
\end{equation}
For a constant $\psi_2$, Eq. (\ref{phi3}) implies
\begin{equation}
 H_0^2 = 4 e^{-4 \psi_2}  .    
\label{h}
\end{equation}
In this case, other equations of motion are summarized by the following
two-dimensional action
\begin{equation}
  I = \frac{1}{2} \int d^2 x \sqrt{- g } e^{- 2 \phi } \Big(
R + 4 ( D \phi )^2 + 4 \Lambda - (D \psi )^2  - \frac{1}{4} 
(D \psi_1 )^2 - \frac{1}{4} e^{-\psi_1} F_2^2 -
 \frac{1}{4} e^{\psi_1} F^2 \Big)  
\label{cghs}
\end{equation}
which can also be obtained from Eq. (\ref{twod}) by plugging Eq. (\ref{h})
in. The constant radius $R_{S^3}$ of the 
three-sphere $S^3$ and the two-dimensional 
cosmological constant $\Lambda$ are related 
to $H_0$ as follows.
\begin{equation}
\Lambda = \frac{1}{R_{S^3}^2} = \frac{2}{|H_0|} = e^{2 \psi_2} 
\label{xxx}
\end{equation}
Now we note that the action Eq.(\ref{cghs}) is precisely that of the
two-dimensional CGHS model coupled with two $U(1)$ gauge fields,
the doubly charged CGHS model.  
We note that the exponential of the
scalar field $\psi_1$ measures the radius of the circle $S^1$.  
We have a manifest $T$-dual invariance; under the $T$-dual 
transformation along the circle $S^1$, that is implemented 
in our context by the transformation $\psi_1 \rightarrow -\psi_1$, 
the winding number of the fundamental string (which produces the 
non-vanishing gauge field strength $F$) gets interchanged
with the Kaluza-Klein momentum number (which produces the non-vanishing
gauge field strength $F$).  As far as the static solutions are concerned, 
the four-torus size, measured by the exponential
of the scalar field $\psi$, becomes a constant if 
we try to avoid naked singularity
solutions as we will show in the next section.  This is consistent 
with the two-dimensional version of the no-scalar-hair theorem.

\section{General Static Solutions}

The classical static equations of motion of the two-dimensional action
Eq. (\ref{cghs}) can be completely solved by resorting to the
method of \cite{klp}.  After the rescaling of the metric 
$g_{\alpha \beta}\rightarrow e^{2\phi} g_{\alpha\beta}$,
the action Eq. (\ref{cghs}) becomes 
\begin{equation}
  I = \frac{1}{2} \int d^2 x \sqrt{- g } e^{- 2 \phi } \Big(
R +  4 \Lambda e^{2\phi} -(D \psi)^2- \frac{1}{4} (D \psi_1 )^2 - 
\frac{1}{4} e^{-\psi_1-2\phi} F_2^2 -
 \frac{1}{4} e^{\psi_1-2\phi} F^2 \Big) .  
\label{a1}
\end{equation}
To follow the method of \cite{klp}, we choose a conformal gauge
\begin{equation}
ds^2=g_{\alpha \beta} dx^{\alpha} dx^{\beta}=-\exp (2\rho)dx^+ dx^- .
\label{a2}
\end{equation}
The equations of motion from the action Eq. (\ref{a1}) are
as follows; by varying Eq. (\ref{a1}) with respect to the two-dimensional
dilaton field $\phi$, we have
\begin{equation}
4\partial_+ \partial_- \rho + 2 \partial_+ \psi \partial_-\psi
+\frac{1}{2}\partial_+ \psi_1 \partial_-
\psi_1 + 2 e^{-\psi_1-2\rho}\Omega F^2_{2+-}
+2e^{\psi_1-2\rho}\Omega F^2_{+-}=0
\label{a3}
\end{equation}
where $\Omega=e^{-2\phi}$.  From the variation with respect to the 
conformal factor $\rho$ of the
metric, we have
\begin{equation}
2\partial_+ \partial_- \Omega +2\Lambda e^{2\rho}-
e^{-\psi_1-2\rho}\Omega^2 F^2_{2+-}
-e^{\psi_1-2\rho}\Omega^2 F^2_{+-}=0 ,
\label{a4}
\end{equation}
and the equations of motion for the scalar fields become
\begin{equation}
\partial_+(\Omega \partial_- \psi)+\partial_-(\Omega \partial_+\psi)=0,
\label{a4-1}
\end{equation}
and 
\begin{equation}
\partial_+(\Omega \partial_- \psi_1)+\partial_-(\Omega \partial_+ \psi_1)
+2e^{-\psi_1-2\rho}\Omega^2 F^2_{2+-}
-2e^{\psi_1-2\rho}\Omega^2 F^2_{+-}=0.
\label{a5}
\end{equation}
We also have the equations for the two $U(1)$ gauge fields
\begin{equation}
\partial_+(e^{-\psi_1-2\rho}\Omega^2 F_{2+-})
=\partial_-(e^{-\psi_1-2\rho}\Omega^2 F_{2+-})=0,
\label{a6}
\end{equation}
and
\begin{equation}
\partial_+(e^{\psi_1-2\rho}\Omega^2 F_{+-})=
\partial_-(e^{\psi_1-2\rho}\Omega^2 F_{+-})=0.
\label{a7}
\end{equation}
The equations of motion, Eqs (\ref{a3})-(\ref{a7}), should be 
supplemented with the gauge constraints resulting from the choice 
of a conformal gauge
\begin{equation}
T_{++}=T_{--}=0,
\label{a8}
\end{equation}
where $T_{\pm \pm}$ is the $\pm \pm$ components of the stress-energy tensor.

Restricting our attention only to the static solutions, we require 
that all fields depend only on a space-like coordinate 
$x \equiv x_+-x_-$.  We also make a
convenient gauge choice for the $U(1)$ gauge fields by introducing two
functions $A(x)$ and $A_2(x)$
\begin{equation}
A_{\pm}=\frac{1}{2}A(x),~~~A_{2\pm}=\frac{1}{2}A_2(x).
\label{a9}
\end{equation}
The static equations of motion are then summarized by a
one-dimensional action
\begin{equation}
I=\int dx \left( 4\Omega' \rho' +2\Lambda e^{2\rho} -2\Omega \psi^{'2}
-\frac{1}{2}\Omega \psi_1^{'2}
+e^{-\psi_1-2\rho}\Omega^2 A_2^{'2}
+e^{\psi_1-2\rho}\Omega^2 A^{'2} \right)
\label{a10}
\end{equation}
where the prime represents the differentiation with respect to 
$x$.
The gauge constraints Eq.(\ref{a8}) become a single condition
\begin{equation}
\Omega''-2\rho' \Omega'+\Omega\psi^{'2}+
\frac{1}{4}\Omega \psi_1^{'2}=0.
\label{a11}
\end{equation}
To integrate the equations from the static action
Eq. (\ref{a10}) under the gauge constraint Eq. (\ref{a11})
for six unknown functions, we
observe that there exist the following six symmetries of the 
action Eq. (\ref{a10})
\begin{eqnarray*}
(a)&&~ \psi \rightarrow \psi + \alpha, \\
(b)&&~ \psi_1 \rightarrow \psi_1 + \alpha,~A_2 \rightarrow 
e^{\alpha/2}A_2,~A \rightarrow e^{-\alpha/2}A, \\
(c)&&~ A_2 \rightarrow A_2 + \alpha, \\
(d)&&~ A \rightarrow A + \alpha, \\
(e)&&~ x \rightarrow x+\alpha,  \\
(f)&&~ x \rightarrow e^{\alpha}x,~\Omega \rightarrow e^{\alpha}\Omega,
~e^{2\rho} \rightarrow e^{-\alpha}e^{2\rho},~A_2 \rightarrow 
e^{-\alpha}A_2,~A \rightarrow e^{-\alpha}A,
\end{eqnarray*}
where $\alpha$ is an arbitrary real parameter of each continuous 
transformation. Therefore, the integration of the second order
differential equations once to get the first order differential
equations is straightforward.  The results are summarized by
the following Noether charge expressions.
\begin{eqnarray}
(a)&&~ \psi_0=\Omega \psi'
\label{a11-1} \\
(b)&&~ \psi_{10}=\Omega \psi_1'-e^{-\psi_1-2\rho}\Omega^2 A_2' A_2
+e^{\psi_1-2\rho}\Omega^2 A' A,
\label{a12} \\
(c)&&~ Q_2=e^{-\psi_1-2\rho}\Omega^2 A_2',
\label{a13} \\
(d)&&~ Q=e^{\psi_1-2\rho}\Omega^2 A',
\label{a14} \\
(e)&&~ c_0=\Omega' \rho' -\frac{1}{2}\Omega\psi^{'2}
-\frac{1}{8}\Omega \psi_1^{'2}-\frac{1}{2}
\Lambda e^{2\rho}+\frac{1}{4}e^{-\psi_1-2\rho}\Omega^2 A_2^{'2}
+\frac{1}{4}e^{\psi_1-2\rho}\Omega^2 A^{'2} ,
\label{a15} \\
(f)&&~s+c_0 x=-\frac{1}{2}\Omega'+\rho' \Omega 
-\frac{1}{2}e^{-\psi_1-2\rho}\Omega^2 A_2' A_2-\frac{1}{2}
e^{\psi_1-2\rho}\Omega^2 A' A.
\label{a16}
\end{eqnarray}
The gauge constraint Eq. (\ref{a11}) and the equation of motion for $\rho$
Eq. (\ref{a4}) give $c_0=0$, thereby reducing the number of constants
of motion from six to five.
The derivation of the solutions is most transparent under the
introduction of a following set of field
redefinitions and a coordinate change
\begin{equation}
\bar{\Omega}=e^{\psi_1/2}\Omega,~e^{2\bar{\rho}}=e^{-\psi_1/2}e^{2\rho},
~d\bar{x}=e^{\psi_1/2}dx.
\label{a17}
\end{equation}
In terms of these redefined field, we can rewrite Eqs. (\ref{a11-1})
-(\ref{a14}) and Eq. (\ref{a16}) as follows.
\begin{eqnarray}
\psi_0&=&\bar{\Omega}\dot{\psi}
\label{a17-1} \\
\psi_{10}&=&\bar{\Omega}\dot{\psi_1}-Q_2 A_2+QA,
\label{a18} \\
Q_2\dot{A}&=&Q e^{-2\psi_1}\dot{A_2},
\label{a19} \\
Q&=&e^{-2\bar{\rho}}\bar{\Omega}^2\dot{A},
\label{a20} \\
\bar{s}&=&-\frac{1}{2}\dot{\bar{\Omega}}+\dot{\bar{\rho}}\bar{\Omega}
-QA,
\label{a21}
\end{eqnarray}
where $\bar{s}=s-\psi_{10}/2$ and the overdot represents 
the differentiation with respect to $\bar{x}$.
Combining Eqs. (\ref{a20}) and (\ref{a21}), we find
\[
(2\bar{s}+2QA)\dot{A}=\frac{d}{d\bar{x}}(Qe^{2\bar{\rho}}
\bar{\Omega}^{-1}),
\]
which, upon integration, becomes
\begin{equation}
Q e^{2\bar{\rho}}\bar{\Omega}^{-1}=QA^2+2\bar{s}A+c \equiv P(A),
\label{a22}
\end{equation}
where we introduce a function $P(A)$ and $c$ is a constant of integration.
Putting Eq. (\ref{a22}) into Eq. (\ref{a20}), we get
\begin{equation}
\bar{\Omega}\frac{dA}{d\bar{x}}=P(A)~~~ \rightarrow ~~~\bar{\Omega}
\frac{d}{d\bar{x}}=P(A)\frac{d}{dA}
\label{a23}
\end{equation}
By changing the differentiation variable from $\bar{x}$ to $A$ with
the help of Eq. (\ref{a23}), we immediately find that Eq. (\ref{a17-1})
can be integrated to yield
\begin{equation}
\psi=\psi_0 \int \frac{dA}{P(A)}+\psi_c
\label{a23-1}
\end{equation}
where $\psi_c$ is the constant of integration. In a similar way,
we can rewrite Eq. (\ref{a18}) and Eq. (\ref{a19}) as
\begin{equation}
P(A)\frac{d\psi_1}{dA}+QA-Q_2A_2=\psi_{10}
\label{a24}
\end{equation}
and
\begin{equation}
Q\frac{dA_2}{dA}=Q_2e^{2\psi_1} .
\label{a25}
\end{equation}
Differentiating Eq. (\ref{a24}) with respect to $A$ and using 
Eq. (\ref{a25}), we get
\begin{equation}
\frac{d}{dA}\left(P(A)\frac{d\psi_1}{dA}\right)+Q-\frac{Q_2^2}{Q}
e^{2\psi_1}=0.
\label{a26}
\end{equation}
By setting
\[
\psi_1=-\frac{1}{2}\ln |P(A)| +\hat{\psi}_1,
\]
and introducing a new variable
\[
\hat{A}=\int \frac{dA}{P(A)}
\]
we find that Eq. (\ref{a26}) is a one-dimensional classical Liouville
equation
\[
\frac{d^2}{d\hat{A}^2}\hat{\psi}_1-\frac{Q_2^2}{|Q|}
e^{2\hat{\psi}_1}=0
\]
which can be exactly solved to give
\begin{equation}
e^{2\psi_1}=\frac{|c_1|}{Q_2^2}\frac{Q}{P(A)}\frac{1}
{\sinh^2 \left[ \sqrt{c_1} \left( \int \frac{dA}{P(A)}+\tilde{c}_1
\right)\right]}
\label{a27}
\end{equation}
where $c_1$ and $\tilde{c}_1$ are constants of integration.
We can find $A_2$ using Eq. (\ref{a24}) and Eq. (\ref{a27})
\begin{equation}
Q_2A_2=-\psi_{10}-\bar{s}-\sqrt{c_1} \coth \left[ \sqrt{c_1}
\left( \int \frac{dA}{P(A)}+\tilde{c}_1\right)\right].
\label{a28}
\end{equation}
We express Eq. (\ref{a15}) in terms of the redefined fields 
Eq. (\ref{a17}), change variable from $x$ to $\hat{A}$, and plug in
Eq. (\ref{a23-1}), Eq. (\ref{a27}) and Eq. (\ref{a28}) to get
\begin{equation}
\left( \frac{d\hat{\phi}}{d\hat{A}} \right)^2 
-\frac{\Lambda}{|Q|}e^{2\hat{\phi}}-D_2 =0
\label{a29}
\end{equation}
where $e^{\hat{\phi}}=\bar{\Omega}|P|^{1/2}$ and 
$D_2=(\psi_0^2+c_1+\bar{s}^2-cQ)/2$.
This equation Eq. (\ref{a29}) is the first integration of the 
one-dimensional classical Liouville equation where $D_2$ plays the 
role of the constant of integration. By integrating Eq. (\ref{a29})
we get
\begin{equation}
\bar{\Omega}^2=\frac{|D_2|}{\Lambda}\frac{Q}{P(A)}
\frac{1}{\sinh^2 \left[ \sqrt{D_2} \left( \int \frac{dA}{P(A)}
+\tilde{c} \right) \right]}
\label{a30}
\end{equation}
where $\tilde{c}$ is a constant of integration.
Finally, we can determine $A$ in terms of $x$ using Eq. (\ref{a23})
\begin{equation}
x-x_0=\int \frac{\Omega}{P(A)} dA.
\label{a31}
\end{equation}

  Eqs. (\ref{a22}), (\ref{a23-1}), (\ref{a27}), (\ref{a28}), 
(\ref{a30}) and (\ref{a31}) are the general static solutions 
of the equations of motion.
We have eleven parameters for the general static solutions of 
the action Eq. (\ref{cghs}). Among possible choices of the
parameters we restrict our attention only to the
case when $c_1>0$, $\bar{s}^2-Qc \ge0$, $c/Q>0$, and $\bar{s}>0$, 
which, as we will see shortly, corresponds to the case considered in 
\cite{hks}.  Under these conditions, we further choose the following 
radial coordinate $r$ and the time coordinate $t$.
\begin{equation}
r^2=-\frac{r_0^2}{2}\left[ \coth (\sqrt{D_2}(\hat{A}(A)+\tilde{c}))-1
\right],~~~~t=x^++x^-
\label{a32}
\end{equation}
where $r_0^2=2\sqrt{D_2Q/c}$. In terms of these $(t,r)$ coordinates,
the ten-dimensional metric in the radial gauge and the dilaton field
can be written as follows.
\begin{eqnarray}
ds^2=&&\frac{f_1(r)}{Z_1(r)}\left[-\beta^2 dt^2 + d\theta^2+f_2(r)
(\beta \cosh \sigma dt + \sinh \sigma d\theta)^2 \right] 
\nonumber \\
&&+\frac{\Lambda^{-1}}{r^2} \frac{dr^2}{Z_0(r)}
+\Lambda^{-1}d\Omega^{(3)}+Z_{\psi}dx^m dx^m ,
\label{a33}
\end{eqnarray}
\begin{equation}
e^{-2\phi^{(10)}}=\frac{\Lambda^{-1}}{r^2}\frac{1}{Z_1(r)}
\label{a34}
\end{equation}
where we introduce five functions
\[ f_1 (r) = \frac{\sqrt{Qc}}{\bar{s} +
   \sqrt{\bar{s}^2 - Qc}} e^{\tilde{c} 
   \sqrt{\bar{s}^2 - Qc}}
   \left( 1 - \frac{r_0^2}{r^2}
   \right)^{(1- \sqrt{(\bar{s}^2 - Qc) / D_2 } ) /2 } ,\]
\[ f_2 (r) = 1 -
   \frac{\bar{s} + \sqrt{\bar{s}^2  - Qc}}
     {\bar{s} - \sqrt{\bar{s}^2  - Qc}}
   e^{ -2 \tilde{c} \sqrt{\bar{s}^2 - Qc}}
   \left( 1- \frac{r_0^2 }{r^2} 
   \right)^{\sqrt{(\bar{s}^2 - Qc) / D_2 } } , \]
\[ Z_0(r) = 1 - \frac{r_0^2}{r^2} , \]
\[ Z_1(r) = \sqrt{\frac{c Q_2^2 }{4 c_1 Q}}
   \left| e^{\sqrt{c_1} ( \tilde{c} - \tilde{c}_1 )}
   \left( 1 - \frac{r_0^2}{r^2} \right)^{(1 - 
   \sqrt{c_1 / D_2 } ) /2 }
   -e^{- \sqrt{c_1} ( \tilde{c} - \tilde{c}_1 )}
   \left( 1 - \frac{r_0^2}{r^2} \right)^{(1 + 
   \sqrt{c_1 / D_2 } ) /2 } \right| , \]
\[ Z_{\psi} = e^{\psi_c-\psi_0\tilde{c}}
   \left( 1 - \frac{r_0^2}{r^2} \right)^{
   \psi_0 / (2 \sqrt{D_2 } ) } , \]
and two parameters
\[ \beta = \sqrt{\frac{c}{4Q}} \ , \  \tanh \sigma = -\frac{c}{|c|}
   \frac{\bar{s} - \sqrt{\bar{s}^2 - Qc}}
     {\sqrt{Qc}} . 
\]
The corresponding two-dimensional metric and two-dimensional
dilaton field are
\begin{equation}
ds^2=-\frac{Z_0(r)}{Z_1(r)Z_2(r)}\beta^2dt^2+\frac{\Lambda^{-1}}{r^2}
\frac{dr^2}{Z_0(r)}
\label{a35}
\end{equation}
\begin{equation}
e^{-2\phi}=\Lambda^{-1/2}r^2 \sqrt{Z_1(r)Z_2(r)}
\label{a36}
\end{equation}
where 
\begin{eqnarray*}
Z_2(r)=&&\sqrt{\frac{Qc}{4(\bar{s}^2-Qc)}} \left|
e^{\tilde{c}\sqrt{\bar{s}^2-Qc}}\left(1-\frac{r_0^2}{r^2}\right)^{
(1-\sqrt{(\bar{s}^2-Qc)/D_2})/2} \right. \\
&&~~~~~~~~~~~~~~~~~~~~~~~~~\left. -e^{-\tilde{c}\sqrt{\bar{s}^2-Qc}}
\left(1-\frac{r_0^2}{r^2}\right)^{(1+\sqrt{(\bar{s}^2-Qc)/D_2})/2}
\right|
\end{eqnarray*}
As was shown in \cite{klp}, we have the black hole solutions when
following two conditions are met
\begin{equation}
\psi_0=0,~~~D_2=c_1~\rightarrow~D_2=c_1=\bar{s}^2-Qc ,
\label{a37}
\end{equation}
and, otherwise, the solutions have a naked singularity.
Therefore, for the black hole solutions, the size of the
four-torus is a fixed value (zero charge) and the scalar 
charge of the circle size field
$\psi_1$ is determined by other constants of motion.
This is basically the no-scalar-hair theorem \cite{klp}.

To see that our black hole solutions indeed represent the solutions
of Ref. \cite{hks}, we make further restrictions on parameters such as
\begin{equation}
 \frac{\bar{s} + \sqrt{\bar{s}^2  - Qc}}
     {\bar{s} - \sqrt{\bar{s}^2  - Qc}} =
   e^{ 2 \tilde{c} \sqrt{\bar{s}^2 - Qc}}
\label{a38}
\end{equation}
\[ 
2  \frac{ |Q_2|}{r_0^2} \sinh ( \sqrt{c_1} 
    ( \tilde{c} - \tilde{c}_1 )  ) = 1 . \]
These two conditions set the radius of the asymptotic circle $S^1$ to be one
and, as shown in Eq. (\ref{a39}), set the asymptotic value of the
potential $A_t$ at the spatial infinity 
to be zero.  In addition, we also set the radii of the 
four-torus
as one by setting $\psi_c=0$, which gives $Z_{\psi}=1$ under the black
hole conditions Eq. (\ref{a37}). Under the conditions Eqs. (\ref{a37})
and (\ref{a38}), we can also verify that $f_1(r)=1$,
$f_2(r)=r_0^2/r^2$ and
\[
Z_1(r)=1+\frac{r_1^2}{r^2},~~~
Z_2(r)=1+\frac{r_2^2}{r^2}
\]
where 
\begin{equation}
 r_1^2=\sqrt{Q_2^2+\frac{r_0^4}{4}}-\frac{r_0^2}{2}
 \ , \ r_2^2=\sqrt{Q^2+\frac{r_0^4}{4}}-\frac{r_0^2}{2}.
\label{chdef}
\end{equation} 
The two $U(1)$ gauge fields become
\begin{equation}
A_t=-\frac{Q\beta}{r^2+r_2^2},
~~~A_{2t}=-\frac{Q_2\beta}{r^2+r_1^2}
\label{a39}
\end{equation}
where we have set $\psi_{10}=2\beta(r_1^2-r_2^2)$, which 
makes the asymptotic value of the gauge field $A_{2t}$ at the
spatial infinity zero.
Of the original eleven constants of integration, we imposed two 
conditions to avoid naked singularities, two conditions
to set the asymptotic value of the potentials to be zero, and 
two conditions to set the asymptotic radii of $S^1$ and $T^4$ 
to be one, resulting $11-6 = 5$ parameters.  Our black hole solutions 
are thus characterized by four parameters $(r_0 , r_1 , r_2 ,
\beta )$ that will be related to the 
two $U(1)$ charges, the mass and the time scale choice.  We note
that the time translational invariance hides one extra parameter.  
The black hole solutions, Eqs. (\ref{a33}), (\ref{a34}) and (\ref{a39}),
are identical to the non-asymptotically flat
solutions of \cite{hks}.

\section{Thermodynamics}

The analysis of the thermodynamics in the two-dimensional dilaton
gravity was performed in Ref. \cite{np}.  In that reference, the 
mass $M$ and the entropy $S$ are given by
\begin{equation}
 M = \frac{1}{2} \left[ e^{-2 \phi} \partial_x g \right]_{\infty}
 \ , \ S = 2 \pi \left[ e^{-2 \phi } \right]_{horizon}
\label{master}
\end{equation}
from their Eq. (2.13)\footnote{We note that our action
Eq. (\ref{cghs}) has a factor $1/2$ in front of it.  In other words,
unlike the convention of Ref. \cite{np} where the
two-dimensional Newton's constant was set to one, the two-dimensional
Newton's constant in our case is two.  Therefore, the mass and
the entropy expressions in Eq. (\ref{master}) are divided by two
compared to Eq. (2.13) of \cite{np}.}
in a gauge where the metric is
$ds^2 = - g dt^2 + g^{-1} dx^2$.  The subscript $\infty$
in the mass formula means that we have to evaluate the expression
at the spatial infinity.  The subscript $horizon$ in the entropy
formula means that we evaluate the expression at the black
hole horizon. 
Under a gauge choice where the metric expression is 
$ds^2 = - g_1 dt^2 + g_2^{-1} dr^2 $, Eq. (\ref{master})
changes into
\begin{equation}
M = \frac{1}{2} \left[ e^{-2 \phi} 
   \sqrt{\frac{g_2}{g_1}} \frac{dg_1}{dr} 
  \right]_{\infty} \ , \
S = 2 \pi  \left[ e^{-2 \phi} \right]_{horizon}  . 
\label{master1}
\end{equation}
Under the same gauge choice, by computing the inverse of
the period of the Euclidean time coordinate for the euclideanized 
metric,
the temperature is given by
\begin{equation}
T = \frac{1}{4 \pi} \left[ \sqrt{\frac{g_2}{g_1}} \frac{dg_1}{dr} 
 \right]_{horizon}
\label{tempfo}
\end{equation}
 As derived in Section III, the metric Eq. (\ref{a35}) 
and the dilaton field Eq. (\ref{a36}) of the two-dimensional 
doubly charged CGHS black holes are
\begin{equation}
ds^2=-\left(1-\frac{r_0^2}{r^2}\right)\left(1+\frac{r_1^2}{r^2}
\right)^{-1} \left(1+\frac{r_2^2}{r^2} \right)^{-1}\beta^2 dt^2
+\frac{\Lambda^{-1}}{r^2}\left(1-\frac{r_0^2}{r^2}\right)^{-1}dr^2
\label{2dmetric}
\end{equation}
\begin{equation}
e^{-2\phi}=\sqrt{\Lambda^{-1}(r^2+r_1^2)(r^2+r_2^2)} .
\label{2dilaton}
\end{equation}
Eqs. (\ref{2dmetric}) and (\ref{2dilaton}) are manifestly symmetric
under the exchange $r_1 \leftrightarrow r_2$ 
(momentum-winding exchange) reflecting the underlying $T$-dual
invariance.
In the case of the doubly charged CGHS black holes, the location
of the horizon is $r = r_0$.  The spatial infinity is $r = \infty$
where the two-dimensional coupling $\exp (\phi )$ vanishes.
Thus, via Eqs. (\ref{master1}) and (\ref{tempfo}), the mass, entropy and 
the temperature of the doubly charged CGHS black holes are computed
to be
\begin{equation}
M = r_0^2 + r_1^2 + r_2^2  ,
\label{mass}
\end{equation}
\begin{equation}
S = 2 \pi \sqrt{\Lambda^{-1}(r_0^2 + r_1^2 ) ( r_0^2 + r_2^2 ) } 
\label{entropy}
\end{equation}
and
\begin{equation}
T = \frac{1}{2 \pi}
  \frac{r_0^2}{\sqrt{\Lambda^{-1}(r_0^2 + r_1^2 ) (r_0^2 + r_2^2 )} } .
\label{tempera}
\end{equation}
Similar to the metric and the dilaton expressions, Eqs. (\ref{mass})
- (\ref{tempera}) are invariant under $r_1 \leftrightarrow r_2$.
Using Eqs. (\ref{mass}) - (\ref{tempera}) and Eq. (\ref{chdef}), we 
compute
\begin{equation}
 dM - T dS  = -  \left[ \beta^{-1} A_t \right]_{horizon} dQ
 - \left[ \beta^{-1} A_{2t} \right]_{horizon}  dQ_2
\label{firstlaw}
\end{equation}
where the value of the electric potentials at the horizon are given
by 
\begin{equation}
 \left[ A_t  \right]_{horizon} = - \frac{Q \beta}{r_0^2 + r_2^2}
\label{pot1}
\end{equation}
\begin{equation}
 \left[ A_{2t} \right]_{horizon} = - \frac{Q_2 \beta}{r_0^2 + r_1^2} .
\label{pot2}
\end{equation}
Eq. (\ref{firstlaw}) is the first law of thermodynamics for
the doubly charged black holes.  We note that we already set 
the value of the electric potentials at the spatial infinity
as zero.  Therefore, the coefficients of $-dQ$ and $-dQ_2$
in Eq. (\ref{firstlaw}) are chemical potentials.  We observe that
the scale factor $\beta$, which is not a gauge-invariant constant
of motion, actually 
drops out in the chemical potentials.     
One interesting observation is that the Noether charges for the
$U(1)$ charges, $Q$ and $Q_2$ in Eqs. (\ref{a13}) and
(\ref{a14}), are the physical charges of the
black holes as shown in the first law of thermodynamics
Eq. (\ref{firstlaw}).  This two-dimensional consideration is
consistent with the IIB supergravity side 
consideration\footnote{Due to our
choice of the constants of the integration, we have to set 
$R=V=\alpha^{\prime} = g = 1 $ in \cite{douglas}.} in 
\cite{douglas} where the RR electric charge is obtained by 
performing the
integration of the gauge field strength over the three-sphere
$S^3$ and it also turns out to be $Q_2$.  In the same
reference, the quantized Kaluza-Klein charge is also  
given by $Q$. 

When there are no $U(1)$ gauge fields, the thermodynamic
quantities we computed in this section reduce to the 
well-known results in the CGHS model \cite{np}, where
the temperature is a strict constant.  With the inclusion of {\em two}
additional $U(1)$ gauge fields, we have an extremal limit
($r_0 \rightarrow 0$ ) where the temperature Eq. (\ref{tempera})
vanishes but the entropy Eq. (\ref{entropy}) becomes finite
\begin{equation}
 S = 2 \pi \Lambda^{-1/2} r_1 r_2 = 2 \pi \sqrt{Q_5 Q_2 Q} , 
\end{equation}
where we used Eq. (\ref{xxx}) to relate the
inverse cosmological constant to the five-brane wrapping
number ($\Lambda = 2/H_0 = 1/ Q_5$ ) \cite{douglas}.  
This is precisely the extremal entropy of the $D1$-$D5$-$KK$
black holes of the IIB supergravity \cite{vafa}.

The mass formula Eq. (\ref{mass}), which is generically
non-extremal, is very simple; it is the
sum of three terms, $r_0^2$, $r_1^2$ and $r_2^2$, reminiscent
of the BPS mass formula for the asymptotically flat 
$D1$-$D5$-$KK$ black holes of the IIB supergravity 
$M_{BPS} = Q + Q_2 + Q_5$.  In the 
extremal ($r_0 \rightarrow 0$) limit, Eq. (\ref{mass}) reduces 
to $M = Q + Q_2 = M_{BPS} - Q_5 $.  The mass $M$,
which was calculated in the two-dimensional framework, actually
represents the mass of the non-asymptotically flat IIB black holes of
\cite{hks}.  The absence of the $Q_5$ contribution to $M$
can be understood if we consider the relationship between
the asymptotically flat and non-asymptotically flat IIB
black holes.  Starting from the asymptotically flat
$D1$-$D5$-$KK$ black holes, we take the $T$-duals that turn
the $D$-fivebranes to the $D$-particles. 
We uplift this ten-dimensional
solutions to the eleven dimensional supergravity and, in this
process, the $D$-particles translate to the gravitons moving
along the spatial $M$ theory circle.  The infinite boost
procedure of Seiberg along the $M$ theory circle maps this
spatial $M$ theory circle toward a (asymptotically) light-like 
circle \cite{hks}.  
Therefore, the original time coordinate gets mapped into
the (asymptotically) light-like time coordinate.  
Thus the energy ($E=p$) of 
gravitons conjugate to the original time gets mapped into the energy
conjugate to the (asymptotically) light-cone time, and this 
new energy vanishes ($E_+ = E-p = 0$).  Upon the dimensional 
reduction along the (asymptotically) light-like circle, we return 
to the ten dimensions and further $T$ duals back turn the $D$-particle 
charges to the $D$-fivebrane charges.  The overall effect of this 
process is
that the original asymptotically-flat $D1$-$D5$-$KK$ black holes
gets mapped into the non-asymptotically flat IIB black holes that we 
analyzed in terms of the doubly charged CGHS black holes (after
an additional $S$-dual transformation). Thus, 
the contribution to the mass $M$ from the fivebranes should be absent.
Indeed, this is what we found from the purely two-dimensional
analysis leading to Eq. (\ref{mass}).  An interesting point is that
the mass itself is conventionally evaluated near the spatial 
infinity (see Eq. ({\ref{master})).  As far as the near-horizon
geometry is concerned, the non-asymptotically flat IIB black holes
and the their dual asymptotically flat IIB black holes are 
indistinguishable.  Therefore, the agreement of the entropy
expression, that is a typical near-horizon quantity, for both cases 
is intuitively clear.  However, since 
two types of black holes have radically different asymptotic
geometries, it is surprising that the mass formulas agree (up to the
aforementioned contribution from the fivebranes) especially from the point
of view of the ten-dimensional physics.  In a similar vein, it was 
observed by Hyun that the greybody factor calculations for 
the doubly charged CGHS black holes and for the $D1$-$D5$-$KK$ 
black holes agree \cite{private}.  Considering the intuitive fact that 
the greybody factor encodes the accumulated contribution from the spatial 
infinity to the black hole horizon, it is also surprising.
 
To better understand the mass formula Eq. (\ref{mass}), we rewrite it
in terms of $Q$ and $Q_2$.
\begin{equation}
 M =  \sqrt{Q_2^2 + \frac{r_0^4}{4}} + \sqrt{Q^2 + \frac{r_0^4}{4}} 
\label{mass2}
\end{equation}
Eq. (\ref{mass2}) formally looks like a threshold bound state
energy of two particles with mass $r_0^2 /2$ with the momentum
$Q$ and $Q_2$, respectively.  In fact, for the massive black 
holes ($r_0^2 \gg Q, Q_2$),
we get 
\[ M \simeq  2 (r_0^2 /2) + \frac{Q_2^2}{2 (r_0^2 /2 )} 
   + \frac{Q^2}{2 (r_0^2 /2 )} + \cdots   \]
while for the extremal black holes where
$r_0$ is vanishingly small, we have $M = Q_2 + Q$,
an ultrarelativistic energy-momentum relation.
In fact, if the string wrapping number $Q_2 = 0$, the natural 
energy produced by a threshold bound state of a particle of 
mass $r_0^2 /2$ at rest and a particle of the same rest mass
moving along the circle $S^1$ with the momentum $Q$ will be
$E = ( r_0^2 /2 ) + \sqrt{Q^2 + \frac{r_0^4}{4}}$.  Now the
$T$-duality between the momentum and the winding will give
Eq. (\ref{mass2}) {\em if} there is no interaction
energy (that will be symmetric under the exchange of $Q$ and
$Q_2$, and vanish when one of them becomes zero), since 
the mass expression should be symmetric under the exchange
of $Q$ and $Q_2$.  Similar to the BPS mass formula where the
supersymmetry ensures the absence of such interaction energy
contribution, the generically non-extremal (thus, non-supersymmetric)
mass expression Eq. (\ref{mass2}) apparently has no interaction
energy contribution.

\section{Discussions}

In this paper, we established the thermodynamics 
for non-asymptotically flat five-dimensional black holes
via the simpler analysis in the two-dimensional system.
This way, by reliably computing the entropy, we were able to 
independently provide the support for the statement of \cite{hks} 
that $D1$-$D5$-$KK$ black holes are $U$-dual to the doubly
charged CGHS black holes.  Furthermore, we showed that
the mass formula computed from the two-dimensional 
perspective reflects the infinite Lorentz boost process along
the $M$ theory circle, which is a part of the $U$-dual 
chain \cite{hks}.

As can be seen from our thermodynamic analysis, the black holes
of the doubly charged CGHS model have rich but fairly simple structures
that capture the essentials of the $D1$-$D5$-$KK$ black
holes of the IIB supergravity.  As such, it provides us with a 
simple toy model to better understand the space-time physics of
string theory.  Intimately related to the doubly
charged CGHS model is the two-dimensional model of McGuigan, 
Nappi, and Yost
\begin{equation}
 I = \frac{1}{2} \int dx^2 \sqrt{-g} e^{-2 \phi} \left(
 R + 4 (D \phi )^2 - \frac{1}{4} \tilde{F}^2 + c \right) 
\label{mnyac}
\end{equation}
which contains a single $U(1)$ gauge field (see also \cite{gibbons}) 
and can be considered as the heterotic string target space
effective action \cite{mny}.  The thermodynamics of this
model was explicitly worked out in \cite{np}, and the resulting
entropy was found to be identical to the $D1$-$D5$-$KK$ black
holes of the IIB supergravity in \cite{teo} when we require that 
the Kaluza-Klein charge and the fundamental string winding number
are the same, thereby effectively reducing the number of
$U(1)$ gauge fields to one.  Starting from the action of the
doubly charged CGHS model, Eq. (\ref{cghs}), we consider the 
case when $\psi_1 = 0$ and $F_1^2 = F^2 = \tilde{F}^2 /2 $.  
The imposition of these
conditions are consistent with the equations of motion, and we
end up recovering the action Eq.(\ref{mnyac}) with $c = 4 \Lambda$.
From this point of view, the McGuigan-Nappi-Yost model can 
also be regarded as the self-dual sector of the doubly
charged CGHS model (or the type IIB supergravity) under the $T$ duality
along the circle where the strings are wrapped. 

By further chains of $U$-dual transformations,
we can map the doubly charged CGHS black holes to the 
three-dimensional charged BTZ black holes \cite{btz},
that is asymptotically
$AdS_3$ \cite{hyun} \cite{welch}.  Considering this fact, the 
doubly charged CGHS model provides yet another way to understand
the physics of the $AdS$ supergravity \cite{ads}.  In fact, a 
qualitative similarity in this regard was noted in \cite{martinec}.
Just like the analysis of the uncharged CGHS model, it 
is sensible to introduce a time-like boundary in our 
two-dimensional model \cite{chung}.  Furthermore, since our model 
is a dimensionally reduced one from the IIB supergravity, we have 
a better understanding of the internal dynamics of the boundary
point.  The dynamical investigation of the doubly charged
CGHS black holes, for example by including the fundamental
string probes, should give us a better understanding of the
$AdS$ supergravity and five-dimensional $D1$-$D5$-$KK$ black holes 
in a much simpler setting of the two-dimensional gravity.
We plan to address this issue in future publication.

\acknowledgements{We were supported in part by the KOSEF grant
971-0201-007-2. This work was supported (in part) by the Korea
Science and Enginerring Foundation (KOSEF) through the Center for
Theoretical Physics (CTP) at Seoul National University (SNU).
Y.K. would like to thank Seungjoon Hyun for useful discussions.}

\end{document}